\documentclass[aps, prl, twocolumn, superscriptaddress]{revtex4-2}

\newcounter{CommentNumber}
\renewcommand{\paragraph}[1]{\stepcounter{CommentNumber}\belowpdfbookmark{#1}{\arabic{CommentNumber}}}
\pdfoutput=1
\usepackage[toc,page]{appendix}
\usepackage[colorlinks]{hyperref}
\usepackage{amsfonts}
\usepackage{amsmath}
\usepackage[utf8]{inputenc}
\usepackage[version=4]{mhchem}
\usepackage{siunitx}
\usepackage{enumerate}
\usepackage{amssymb}
\usepackage{layouts}
\usepackage[normalem]{ulem}

\newcommand{\upperRomannumeral}[1]{\uppercase\expandafter{\romannumeral#1}}
\DeclareMathOperator{\sinc}{sinc}
\renewcommand{\Im}{\textrm{Im}}
\begin{document}

\title{Aharonov-Bohm magnetism in open Fermi surfaces}
\author{Kostas Vilkelis}
\affiliation{Qutech, Delft University of Technology, Delft 2600 GA, The Netherlands}
\affiliation{Kavli Institute of Nanoscience, Delft University of Technology, Delft 2600 GA, The Netherlands}
\author{Ady Stern}
\affiliation{Weizmann Institute of Science, Rehovot, 76100, Israel}
\author{Anton Akhmerov}
\affiliation{Kavli Institute of Nanoscience, Delft University of Technology, Delft 2600 GA, The Netherlands}

\date{\today}

\begin{abstract}
Orbital diamagnetism requires closed orbits according to the Liftshiftz-Kosevich theory. Therefore, one might expect that open Fermi surfaces do not have a diamagnetic response. Contrary to this expectation, we show that open orbits in finite systems do contribute a magnetic response which oscillates between diamagnetism and paramagnetism. The oscillations are similar to the Aharonov-Bohm effect, because the oscillation phase is set by the number of flux quanta through the area defined by the width of the sample and the distance between adjacent atomic layers. The magnetic response originates from the closed trajectories formed by counter-propagating open orbits coupled via specular boundary reflections. The phenomenon acts as a probe of the phase coherence of open electron trajectories.
\end{abstract}

\maketitle

\section{Introduction}

\paragraph{Diamagnetism is a quantum property of closed orbits}

According to the classical theory by Langevin, diamagnetism is a result of the cyclotron motion of electrons in a magnetic field~\cite{langevin1905theorie}.
While this explanation provides an intuitive picture, it is incorrect due to the Bohr--van Leeuwen theorem~\cite{Savoie_2015, bohr-dissertation} that proves the absence of magnetic response in classical mechanics.
On the other hand, a more modern interpretation by Liftshiftz-Kosevich~\cite{lift_kos} explains diamagnetism as a result of quantized closed orbits along the Fermi surface.
The picture by Liftshiftz-Kosevich is simple yet incredibly successful at explaining phenomena like de Haas-van Alphen (dHVA) diamagnetic oscillations~\cite{haas} through the Fermi surface shape of metallic systems~\cite{gold_fs, alkali_fs}.

\paragraph{Orbits that do not close cannot be quantized and thus are not magnetic}

\begin{figure}[tbh]
  \includegraphics{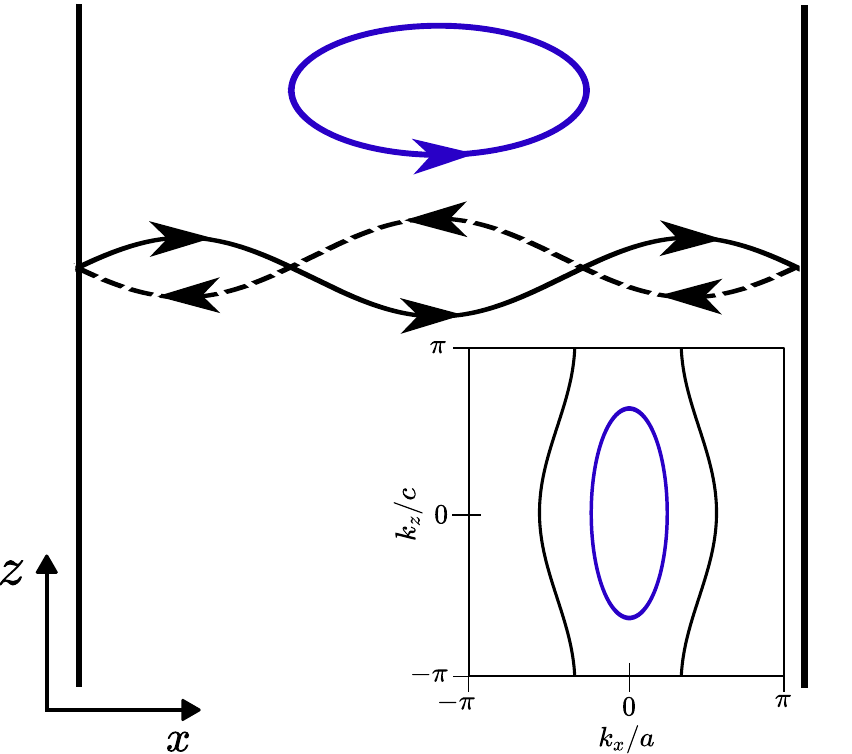}
  \caption{An example of closed (blue curve) and open (black curve) orbits. The right-moving (solid black line) and the left-moving (dashed black line) open trajectories are connected through boundary reflections. The inset at the bottom right shows the corresponding Fermi surface which is closed(open) for the blue(black) curve.}
  \label{fig:1}
\end{figure}

Because ballistic orbits in the magnetic field are rotated and rescaled cuts of the Fermi surface, a Fermi surface that spans the whole Brillouin zone results in an open cyclotron orbit.
An example of an open orbit is shown in Fig.~\ref{fig:1} by the black curve.
These orbits appear in metals such as copper~\cite{copperfs} and gallium~\cite{galliumFS} or in highly anisotropic materials like delafossites~\cite{dela_fs}.
The Liftshiftz-Kosevich theory~\cite{lift_kos} states that open orbits do not have a magnetic response.
However, in multi-band materials with magnetic breakdown regions, it is possible to couple several open orbits into an effective closed orbit~\cite{KAGANOV1983189}.
Such effective closed orbits have a magnetic response, but the contribution is exponentially small.
On the other hand, open orbits in single-band materials do not close.
As a result, the open orbits cannot be quantized and thus do not have a magnetic response according to the Bohr--van Leeuwen theorem~\cite{Savoie_2015, bohr-dissertation}.
That raises the question of whether it is possible to observe quantum interference phenomena in open orbits without magnetic breakdown.

\paragraph{Open orbits connected via boundary reflections lead to Aharonov-Bohm magnetic oscillations}

In this paper, we develop a theory of the orbital magnetic response of open orbits in finite samples and predict magnetic oscillations alternating between diamagnetism and paramagnetism.
Similar to the $h/e$ oscillations of magnetoresistance oscillations in layered materials~\cite{he_osc}, these oscillations have the frequency of the Aharonov-Bohm effect~\cite{ab} through the loop defined between the adjacent conducting atomic layers and the width of the sample.
In addition to requiring ballistic phase-coherent propagation, we find that these magnetic oscillations are sensitive to boundary quality: diffusive boundaries destroy the effect.
With these conditions fulfilled, we predict that this phenomenon has a strength comparable to Landau diamagnetism.

\section{Open Orbit Quantization via Boundary Reflections}

\paragraph{We model an open Fermi surface by a quasi-dimensional system}

We begin from considering an open Fermi surface in layered materials with weak interlayer coupling, however our theory equally applies to any other open Fermi surfaces.
The dispersion of such a layered system is
\begin{equation}
  \varepsilon{(\kappa_x, k_z)} = \varepsilon_\parallel{(\kappa_x)} + 2 t_\perp \cos{\left(k_z c\right)},
  \label{eq:dispersion}
\end{equation}
where $c$ is unit cell spacing along the $z$-direction, $t_\perp$ is the interlayer coupling and $\varepsilon_\parallel{(k_\kappa)}$ is a general dispersion. 
For brevity, we omit the $y$-dimension here and will introduce it later on.
We linearize the dependence of $\epsilon_\parallel$ on $\kappa_x$ at energy $E$:
\begin{equation}
  \begin{aligned}
  & \varepsilon_\parallel{(\kappa_x)}  \approx   E + \hbar v_x(E) \left[\kappa_x - k_x(E)\right] \\
  & \varepsilon_\parallel(k_x) = E, \quad \hbar v_x(E) =  \frac{\partial \varepsilon_\parallel(k_x(E))}{\partial k_x},
  \label{eq:dispersion_lin}
\end{aligned}
\end{equation}
where $k_x(E)$ and $v_x(E)$ are momentum and and velocity along x-direction at energy $E$ when  $k_z = \pi/(2c)$, such that the out-of-plane energy is zero.
Note that we require Eq.~\eqref{eq:dispersion} to be open at Fermi level $\varepsilon=\mu$ along the $k_z$ direction.

\paragraph{The addition of in-plane magnetic field introduces open trajectories}

The addition of a homogeneous magnetic field perpendicular to the interlayer coupling $t_\perp$ introduces open orbits that run along the open direction of the Fermi surface.
The magnetic field points along the $y$-direction $\mathbf{B} = (0, B, 0)$, and introduces a vector potential $\mathbf{A} = (0, 0, -Bx)$ in the Landau gauge.
It enters the Hamiltonian Eq.~\eqref{eq:dispersion} via the Peierls substitution~\cite{peierls1933theorie} $k_z \to k_z - \frac{e}{\hbar} B x$.
In this case, $\kappa_x$ is not conserved anymore and therefore we substitute Eq.~\eqref{eq:dispersion_lin} into Eq.~\eqref{eq:dispersion} at fixed energy $\varepsilon(\kappa_x, k_z)= E$ in order to define the local momentum along $x$:
\begin{equation}
  \kappa_x(E, k_z, x) = \pm \left[k_x (E)- \frac{2 t_\perp}{\hbar v_x(E)} \cos{\left(k_z c-\frac{e}{\hbar} c B x\right)}\right]
  \label{eq:kx}
\end{equation}
Whenever $k_x(E) > 2 t_\perp/(\hbar v_x(E))$, the trajectory in Eq.~\eqref{eq:kx} is open because $\kappa_x$ stays strictly positive/negative.

\paragraph{Boundary reflections quantize the open orbits}

The semiclassical motion in a magnetic field follows constant energy lines in momentum space and a real-space trajectory that is perpendicular to the momentum-space one. For the open Fermi surface we consider, this implies periodic motion in the $z$-direction, but not in the $x$-direction. In the latter, the electron flips its direction of motion only when it scatters off a boundary. For specular scattering off the boundary, the allowed trajectories are those that fulfil by the Bohr-Sommerfeld quantization rule~\cite{Alexandradinata_2018} given via WKB theory~\cite{Bender1999advanced}:
\begin{equation}
  S (E, k_z) = \oint \kappa_x (E, k_z) d x = 2 \pi \left(n+\gamma\right),
  \label{eq:bohrsommer}
\end{equation}
where $n \in \mathbb{Z}$ and $\gamma$ is the Maslov index $\gamma=1/2$ for soft potential turning points and $\gamma=0$ for hard-wall boundaries.
To calculate the quantized spectrum, we substitute Eq.~\eqref{eq:kx} into Eq.~\eqref{eq:bohrsommer}:
\begin{gather}
  S (E, k_z) = 2 k_x(E) W - \frac{2\Gamma(\phi)\cos{\left(k_z c\right)}}{\Delta E_x}, \\
  \Delta{E_x}(E) = \frac{\hbar v_x(E)}{W}, \quad \phi = \frac{e}{\hbar}cWB, \quad \Gamma = 2 t_\perp  \sinc{\left(\phi/2\right)},
\label{eq:action}
\end{gather}
with $\Delta{E_x}$ the energy spacing along $x$, $\phi$ is the number of magnetic flux quanta (in units of $2\pi$) passing through the loop of area $cW$.
The above equation for action is quantized in terms of $2\pi n$ ($\gamma = 0$ for hard-wall boundaries).

\paragraph{The energy bandwidth of the out-of-plane states oscillates and decays with magnetic field}

\begin{figure}[h!]
  \includegraphics{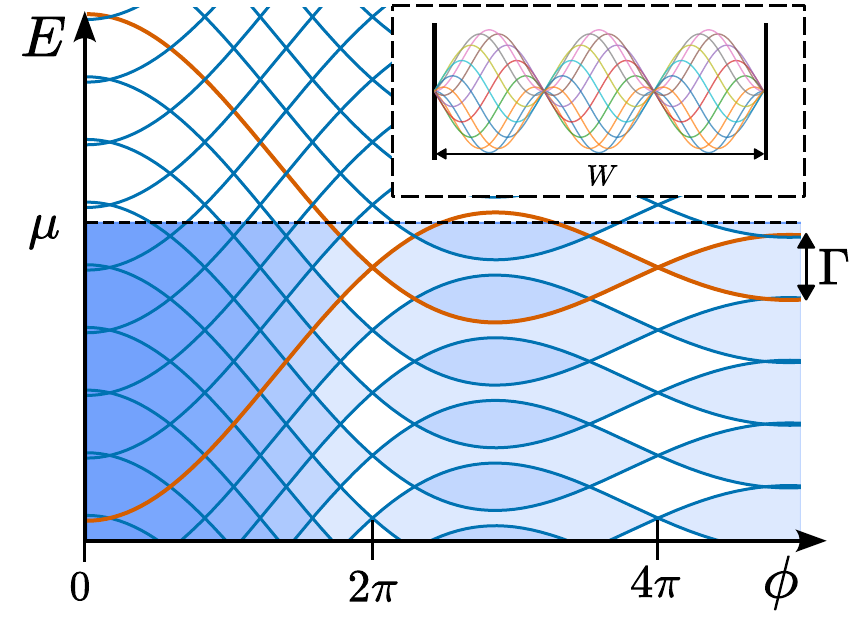}
  \caption{Plot of multiple displaced and overlapping $k_z$ bands (blue curves) as a function of flux quanta $\phi$ passing through the system. The orange curve highlights one such band and its variation of bandwidth $\Gamma$ with $\phi$. The blue filling illustrates the occupation of bands below the chemical potential $\mu$ (dashed line), with the intensity indicating the number of overlapping bands at that point. As the bandwidth changes with respect to $\mu$, the occupation of the bands changes which leads to a magnetic response. When an integer number of flux quanta pass through the system, different $k_z$ trajectories are identical since they lead to the same energy as shown in the top right inset and therefore the bandwidth collapses.}
  \label{fig:2}
\end{figure}

The Bohr-Sommerfeld quantization in Eq.~\eqref{eq:action} defines the allowed energies by the relation:
\begin{equation}
  \hbar v_x{(E_n)} \left[\frac{\pi n}{W}-k_x(E_n) \right] + \Gamma \cos{\left(k_z c\right)} = 0.
  \label{eq:B_dispersion_problem}
\end{equation}
The solution to Eq.~\eqref{eq:B_dispersion_problem} is
\begin{equation}
  E_n = \varepsilon_\parallel{\left(\frac{\pi n}{W}\right)} + \Gamma \cos{\left(k_z c \right)},
  \label{eq:B_dispersion}
\end{equation}
where one identifies $\Gamma$ as the effective bandwidth of $k_z$ band as defined in Eq.~\eqref{eq:action}.
The bandwidth oscillates with the number of flux quanta $\phi$ threading a rectangle of size $Wc$. 
The oscillations decay in a Fraunhoffer-type way, and their periodicity is that of the Aharonov-Bohm effect,  as shown in Fig.~\ref{fig:2}.
Furthermore, when the number of flux quanta is an integer, the bandwidth $\Gamma$ collapses to zero in which case the different $k_z$ channels decouple.
In the opposite limit, we see that if we take $B = 0$ in Eq.~\eqref{eq:B_dispersion}, we recover the original dispersion given by Eq.~\eqref{eq:dispersion}.

\paragraph{}

\section{Diamagnetic Response of an Open Fermi surface}

\paragraph{To calculate magnetisation, we apply analytic continuation to the density of states expressed through the action of a trajectory}
To find the total magnetisation of the system, we reintroduce back the $y$-dimension.
We start with a zero-temperature case and consider finite temperature later.
In this case, the magnetisation of the system at fixed $k_y$ is:
\begin{equation}
  M(\mu, k_y) = \frac{d}{dB} \int_{-\infty}^{\mu}  E \rho(E, k_y) dE,
\label{eq:mag_defin}
\end{equation}
where $\rho(E, k_y)$ is the density of states along $x$ at energy $E$ and wavevector $k_y$.
We express the density of states $\rho$ through the action of a trajectory in Eq.~\eqref{eq:action}, similar to the work by Doron and Smilansky~\cite{dosaction}:
\begin{equation}
  \rho(E, k_y) = -\frac{1}{W \pi^2} \int_{-\pi/c}^{\pi/c} dk_z \frac{d}{dE} \Im \ln{\left(1-e^{iS \left(E+i0^+\right)}\right)},
  \label{eq:dos}
\end{equation}
with $\Im$ the imaginary part.
The integral in Eq.~\eqref{eq:mag_defin} is difficult to compute because it contains a highly oscillatory integrand along energy $E$ given by Eq.~\eqref{eq:dos}.
Therefore we perform analytic continuation of Eq.~\eqref{eq:mag_defin} into the complex energy $E + i\mathcal{E}$ (see supplementary).
The analytic continuation converts the highly oscillatory terms in Eq.~\eqref{eq:dos} into exponentially decaying away from $\mathcal{E} = 0$ and fixes the convergence of the integral.
That allows us to linearise the dispersion in Eq.~\eqref{eq:dispersion_lin} around the Fermi level $\mu$ and compute the magnetisation at $k_y$:
\begin{equation}
\begin{aligned}
    M(k_y, \mu) & = \frac{1}{W c \pi} \frac{d \Gamma}{dB} \sum_{n=1}^{\infty} \frac{\sin{\left(2n k_F W\right)}}{n} J_1\left(\frac{2n\Gamma} {\Delta{E_x}}\right), \\
    k_F(k_y) & = k_x(\mu, k_y), \quad \Delta{E_x}(k_y) = \frac{\hbar v_F(k_y)}{W}, \\ v_F(k_y) & = v_x(E, k_y), 
  \label{eq:mag_full}
\end{aligned}
\end{equation}
where $J_1$ is a Bessel function of the first kind and $k_F$, $v_F$ and $\Delta{E_x}$ are the Fermi momentum, velocity and energy spacing along the x-direction. For $k_y=0$, Eq.~\eqref{eq:mag_full} is the magnetisation of a 2D system with an open Fermi surface.

\paragraph{The magnetisation oscillates with the flux, energy spacing and chemical potential}

Equation~\eqref{eq:mag_full} presents three types of oscillations. The first is the oscillatory Aharonov-Bohm dependence of $\Gamma$ on the flux $\phi$.  The second is oscillations with $\Gamma/\Delta E_x$ that originate from commensuration of the energy separation $\Delta E_x$ between quantized states in the $x$-direction with the bandwidth $\Gamma$. These two types depend on the flux $\phi$.  The third type is oscillations with $k_FW$ that originate from the position of the chemical potential with respect to the centre of a $k_z$ band.
The sum over $n$ represents the different Fourier components of the oscillations with respect to the chemical potential $\mu$.

\paragraph{We integrate out all possible trajectories via the steepest-descent approximation to find 3D magnetisation}

In 3D, the total magnetisation per unit volume is:
\begin{equation}
  \mathcal{M}(\mu) = \frac{1}{\pi}\int_{FS} M(k_y, \mu) dk_y,
  \label{eq:3D_integral}
\end{equation}
where the integral over $k_y$ is along the Fermi surface.
We utilize the steepest-descent method to evaluate the leading order contributions to this integral originating from its behavior near the maxima of $k_F(k_y)$.
To do so, we define the maxima of the Fermi wavevector along $x$ as $K_F$ and the corresponding Fermi surface curvature at these points:
\begin{equation}
\begin{aligned}
  K_F = k_F(k_{y,0}), \quad \frac{dk_F(k_{y,0})}{dk_y} = 0, \\
  -\frac{d^2k_F(k_{y,0})}{d^2 k_y} = -\frac{\partial^2 \varepsilon_\parallel}{\partial k_y^2} \frac{\partial k_F}{\partial \varepsilon_\parallel} = \left(\frac{m_y V_F}{\hbar} \right)^{-1} > 0,
  \label{eq:FS_curvature}
\end{aligned}
\end{equation}
where $V_F = v_F(k_{y,0})$ and $\Delta{E_x} = \Delta{E_x}(k_{y,0})$ are the Fermi velocity and energy spacing along x-direction at $K_F$.
We substitute Eq.~\eqref{eq:FS_curvature} into Eq.~\eqref{eq:3D_integral}, deform the integration contour along the steepest descent and obtain total magnetisation:
\begin{equation}
  \begin{aligned} 
    \mathcal{M}(\mu) & = \mathcal{M}_0 \frac{d {\rm sinc(\phi/2)}}{d\phi} \sum_{n=1}^{\infty} \frac{\sin{\left(2n K_F W \right)}}{n^{3/2}} J_1\left(\frac{2n\Gamma} {\Delta{E_x}}\right), \\
    \mathcal{M}_0  & = \frac{2 e t_\perp}{W \pi^{3/2}} \sqrt{k_{y, \textrm{eff}} W}, \quad k_{y, \textrm{eff}} = \frac{m_y V_F}{\hbar},
  \label{eq:3D_integral_main}
\end{aligned}
\end{equation}
where $k_{y, \textrm{eff}}$ is the effective Fermi $y$-momentum below which all the trajectories are orientated along the $x$-direction.
\begin{figure}[h!]
  \includegraphics{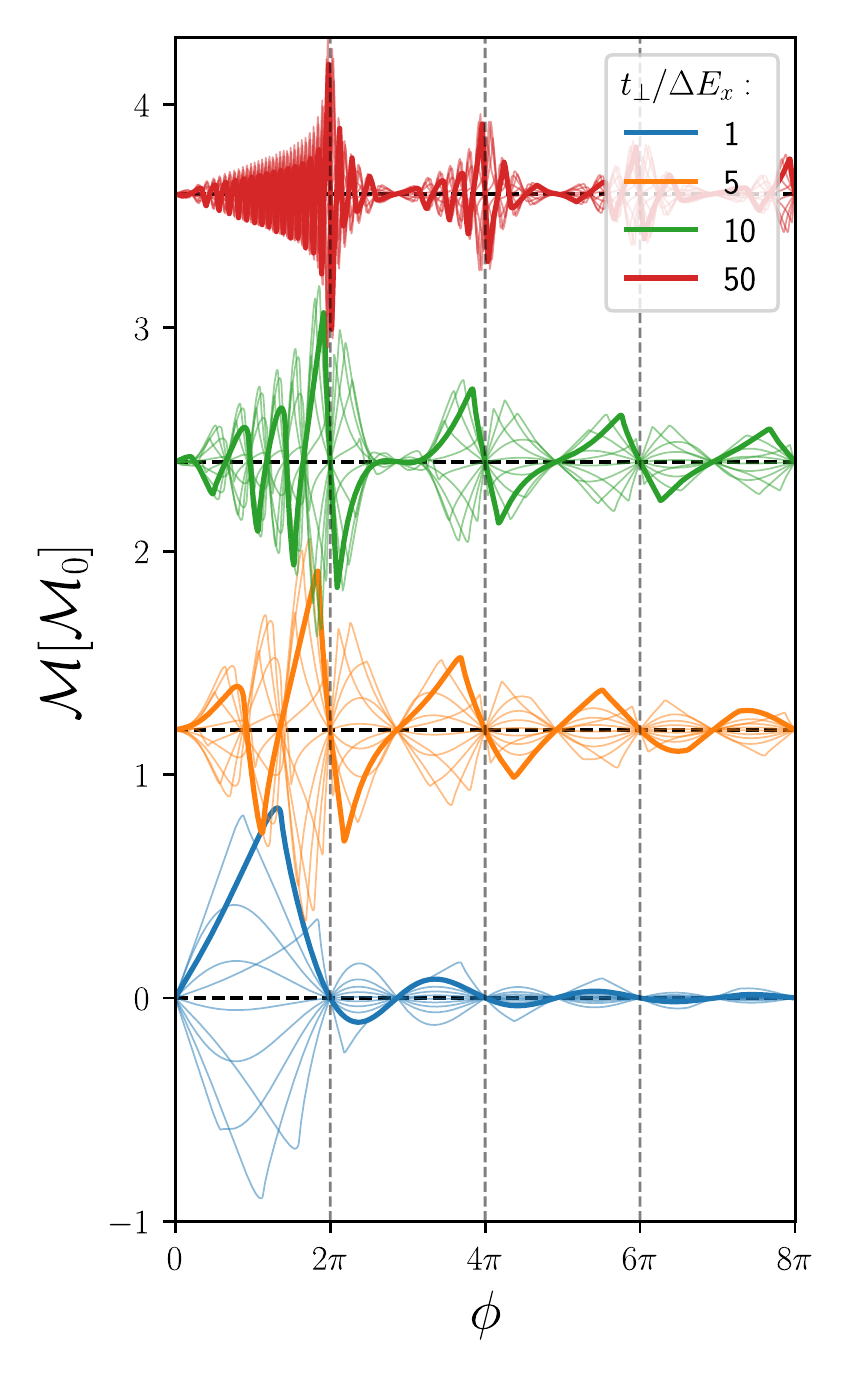}
  \caption{The magnetisation as a function of magnetic flux quanta passing through the system for different $t_\perp / \Delta{E_x}$ ratios. The main (thick) curves are evaluated at $K_F W = \pi/8$ whereas the thin secondary (thin) curves are evaluated at other $K_F W$  values.}
  \label{fig:3}
\end{figure}

\paragraph{Thermal broadening smooths out high-frequency components of magnetisation}

The magnetisation in Eq.~\eqref{eq:3D_integral_main} is a complex oscillatory function of the Fermi wavevector and the magnetic field $B$.
To simplify the expression, we consider thermal broadening of the order of miniband spacing, $k_B T \approx \Delta{E_x}$, which suppresses the terms with $n>1$ (see the supplementary material for details).
\begin{equation}
\begin{aligned}
  \mathcal{M}(\mu, k_B T \approx  \Delta{E_x}) \approx &  \\ \mathcal{M}_0 \frac{d {\rm sinc(\phi/2)}}{d\phi} \sin{\left(2 K_F W \right)} J_1\left(\frac{2\Gamma} {\Delta{E_x}}\right).
 \end{aligned}
  \label{eq:3D_integral_main_T}
\end{equation}
In Eq.~\eqref{eq:3D_integral_main_T}, there are two distinct regimes: the narrow sample, where $t_\perp / \Delta{E_x} \ll 1$ and the wide sample, where $t_\perp / \Delta{E_x} \gg 1$.
These two limits correspond to the presence of either a single $k_z$ energy band (given by Eq.~\eqref{eq:B_dispersion}) at the Fermi level or multiple overlapping bands.

\paragraph{In the single band limit, the magnetisation displays Aharonov-Bohm type oscillations}

In the narrow sample limit $t_\perp / \Delta{E_x} \ll 1$, we expand the Bessel function $J_1$ in Eq.~\eqref{eq:3D_integral_main_T} for small arguments and find a simplified form of magnetisation:
\begin{equation}
\begin{aligned}
  \mathcal{M}(\mu, k_B T \approx \Delta{E_x}) \approx &\\
  -\frac{2\mathcal{M}_0}{\phi^3} \frac{t_\perp}{\Delta{E_x}}&\sin{\left(2 K_F W \right)} \left(2- \phi \sin{\phi} - 2 \cos{\phi}\right).
  \label{eq:mag_single}
\end{aligned}
\end{equation}
The magnetisation in Eq.~\eqref{eq:mag_single} oscillates with $\phi$, the number of flux quanta passing through an area $cW$ similar to the Aharonov-Bohm effect, however the oscillations decay with $\phi^3$.
Figure~\ref{fig:2} provides a qualitative explanation of this behavior as a response of a partially occupied band with bandwidth that both oscillates and decays with $\phi$.
Thes oscillations are distinct from the dHVA diamagnetism~\cite{haas} that oscillates with inverse magnetic field $1/B$ because the cyclotron orbit shrinks with magnetic field.
Due to their similarity with the Aharonov-Bohm effect, we name the magnetisation oscillations of open Fermi surfaces Aharonov-Bohm magnetism.

\paragraph{In the overlapping band limit, the magnetisation reaches maxima for an integer number of flux quanta passing through $cW$}

In the wide sample limit $t_\perp / \Delta{E_x} \gg 1$, Eq.~\eqref{eq:3D_integral_main_T} exhibits a combination of multiple frequency oscillations combined with an overall decay, as shown in Fig~\ref{fig:3}.
However, Aharonov-Bohm magnetism is still evident in this regime because regardless of the chemical potential, the amplitude of the magnetization oscillations reaches its maximum whenever an integer number of flux quanta passes through the area $cW$.

\section{Practical Considerations}

\paragraph{Aharonov-Bohm diamagnetism magnitude is related to mass anisotropy and sample width}

In order to estimate the magnitude of the magnetic susceptibility $\chi$, we expand the magnetisation in Eq.~\eqref{eq:3D_integral_main_T} to first order in flux $\phi$:
\begin{equation}
  \begin{aligned}
  \chi = \mu_0 \frac{d\mathcal{M}}{dB} \approx - \mu_0 \frac{\mathcal{M}_0}{12}\frac{d \phi}{dB} \sin{\left(2 K_F W \right)} J_1\left(\frac{4 t_\perp}{\Delta{E_x}}\right),
  \label{eq:ab_suscept}
  \end{aligned}
\end{equation}
where $\mu_0$ is the vacuum permeability.
To make the interpretation clearer we consider Landau diamagnetism~\cite{landaudia} of an isotropic dispersion ($m_y/m_x = 1$) and Fermi wavevector $K_F$:
\begin{equation}
  \chi_{L} = - \mu_0 \frac{e^2 K_F}{12 \pi^2 m_\parallel}.
  \label{eq:landau}
\end{equation}
The ratios between Aharonov-Bohm diamagnetism in Eq.~\eqref{eq:ab_suscept} and Landau diamagnetism in Eq.\eqref{eq:landau} in the narrow and wide sample limits are:
\begin{equation}
\begin{aligned}
& \frac{\max(\chi)}{\chi_L}  \approx \\
& \begin{cases}
 \sqrt{\frac{m_\parallel}{m_\perp}} \left(K_F W\right)^{-1/2} \sqrt{k_{y, \textrm{eff}} W} & 4 t_\perp \gg \Delta{E_x} \\
\sqrt{\pi} \left(\frac{m_\parallel}{m_\perp}\right)^2 \left(\frac{W}{c}\right)^3 \left(K_F W\right)^{-3/2} \sqrt{k_{y, \textrm{eff}} W} &4 t_\perp \ll \Delta{E_x}
\end{cases}
\end{aligned}
\label{eq:susc_OOM}
\end{equation}
where we substituted $t_{\perp} = \hbar^2/(2 m_{\perp} c^2)$ and $V_F = \hbar K_F/m_\parallel$ where $m_\perp$ is the mass along the $z$ direction, $m_\parallel$ is the mass along the in-plane direction.
We see from Eq.~\eqref{eq:susc_OOM} that in both limits AB diamagnetism favors large mass anisotropy $m_\parallel/m_\perp$ and flat in-plane Fermi surfaces that maximize $k_{y, \textrm{eff}}$.
However, the narrow sample limit susceptibility scales much better with mass anisotropy and further scales with the width of the sample $W$.
As a check, we use typical parameters of microscopic delafossite samples\cite{delafossites_review}, and observe that a sample with Fermi momentum $K_F W \approx 10^{4}$ and $\sqrt{k_{y, \textrm{eff}} W} \approx 1$, mass anisotropy $m_\parallel/m_\perp = 10^{-3}$, lattice spacing $c = \SI{1}{\angstrom}$ and the width of the sample $W = \SI{1}{\micro\metre}$ generates diamagnetism of the same order as Landau diamagnetism $\chi/\chi_L \approx 1$.

\paragraph{Diffusive boundaries and bulk scattering destroy the Aharonov-Bohm diamagnetic oscillations}

The Bohr-Sommerfeld quantization condition in Eq. \eqref{eq:bohrsommer} relies on specular boundary reflections at the ends of the sample to close the trajectory.
To examine the role of diffusive boundary scattering with the specular reflection probability equal to $r$, we evaluate the magnetisation numerically in finite samples.
We observe that the amplitude of the magnetisation is proportional to $r^2$, consistent with the closed trajectory requiring two specular reflections.
Furthermore, we remark that random bulk scattering and dephasing work in the same way as diffusive boundary reflection: the probability to encounter a random scattering/dephasing event in a $2W$ width sample with mean-free path/phase coherence length $l_{0/\phi}$ is $exp(-2W/l_{0/\phi})$.
Therefore, the strength of Aharonov-Bohm magnetism depends on both mean-free-path and boundary quality:
\begin{equation}
\mathcal{M} \propto r^2 \left[1-\exp\left(-\frac{l_{0/\phi}}{2W}\right)\right].
\label{eq:mag_diffusive}
\end{equation}

\paragraph{Necessary conditions to observe Aharonov-Bohm magnetism}

Finally, we summarize the necessary conditions required to observe Aharonov-Bohm magnetism:
\begin{enumerate}
  \item Open component to the Fermi surface.
  \item Phase coherence length and mean-free path larger than the sample width, $W \leq l_\phi, l_0$.
  \item High-quality sample boundaries to ensure specular reflections.
\end{enumerate}
One candidate family of materials which fulfill conditions 1. and 2. are the delafossites~\cite{delafossites_review} like \ce{PdCoO_2} and \ce{PtCoO_2}.
Delafossites are highly anisotropic materials with a cylindrical Fermi surface~\cite{dela_fs} and mean-free path on the order of $\SI{20}{\micro\meter}$~\cite{delafossite_mfp}.
Additionally, the hexagonal Fermi surface in delafossites allows one to align a sample in a way that does not permit trajectories along the magnetic field direction and thus maximizes $k_{y, \textrm{eff}}$.
An alternative candidate material is elemental copper~\cite{copperfs}.
Despite not having a fully open Fermi surface, it does have small open components.
Even though that reduces the number of possible open trajectories (and thus $k_{y, \textrm{eff}}$), the out-of-plane mass $m_\perp$ in copper is smaller and thus more favourable than in delafossites.
Additionally, it is possible to engineer copper samples with a mean-free path well into the micrometre scale~\cite{copper_mfp}.
However, in both cases, the sample boundaries pose a significant bottleneck which should be overcome for the effect to be observed.

\begin{acknowledgments}
AS thanks the Israeli Science Foundation Quantum Science and Technology grant no. 2074/19, the CRC 183 of the Deutsche Forschungsgemeinschaft for funding. 
The project received funding from the European Research Council (ERC) under the European Union’s Horizon 2020 research and innovation program grant agreements No. 788715 (LEGOTOP) and No. 828948 (AndQC).
The work was also supported by the NWO VIDI Grant (016.Vidi.189.180).
\end{acknowledgments}

A.S. formulated the initial project idea.
All authors derived the theory.
K.V. ran numerical calculations to verify the theory with input from A.A.
K.V. authored the manuscript with input from other authors.

\bibliography{paper.bib}

\pagebreak
\widetext
\begin{center}
\textbf{\large Supplemental Materials}
\end{center}
\setcounter{equation}{0}
\setcounter{figure}{0}
\setcounter{table}{0}
\setcounter{page}{1}
\makeatletter
\renewcommand{\theequation}{S.\arabic{equation}}
\setcounter{figure}{0}
\renewcommand{\figurename}{Fig.}
\renewcommand{\thefigure}{S.\arabic{figure}}
\title{Supplementary material}

\section{Diamagnetic Response Derivation}
\subsection{2D Result}
\label{appendix:2derivation}

\begin{figure}[tbh]
  \includegraphics{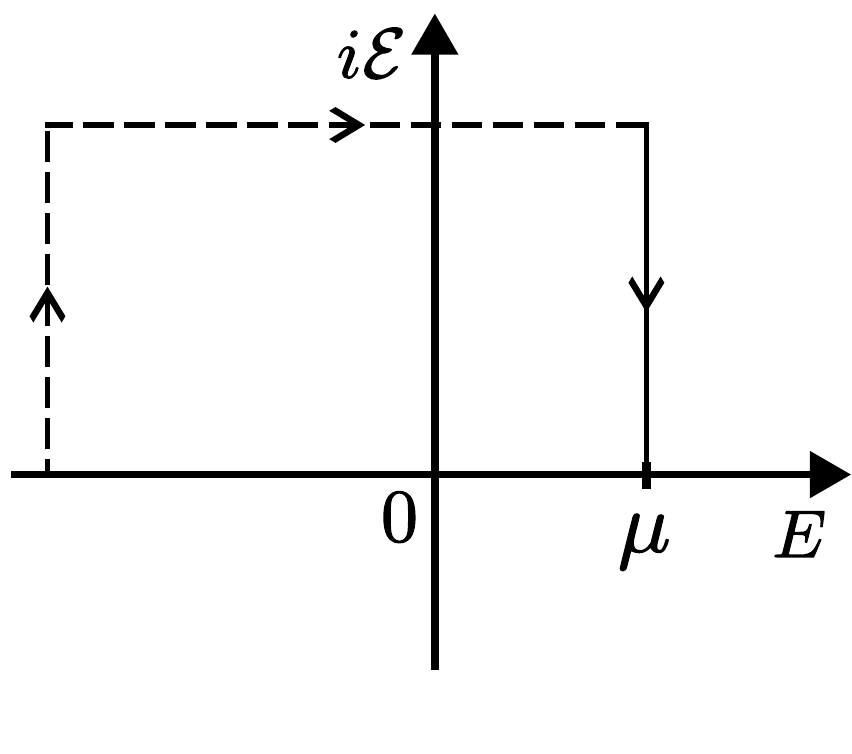}
  \caption{Integration contour along real energy deformed into the complex plane through analytical continuation. The dashed parts of the contour are chosen sufficiently far away such that they do not contribute to the overall integral and the only contribution comes from the solid vertical contour line.}
  \label{fig:suppl}
\end{figure}

We substitute the density of states in Eq.~\eqref{eq:dos} into the definition of magnetisation in Eq.~\eqref{eq:mag_defin}:
\begin{equation}
  \begin{aligned}
  M(k_y, \mu) =
  &-\frac{\Im}{W \pi^2} \frac{d}{dB} \int\limits_{-\infty}^{\mu}  \int\limits_{-\pi}^{\pi} dE dk_z E \frac{d}{dE} \ln{\left(1-e^{iS \left(E+i0^+\right)}\right)} = \\
  & -\frac{1}{W \pi^2} \int_{-\pi/c}^{\pi/c} dk_z \Im  \frac{d}{dB} \biggl(\mu \ln{\left(1-e^{iS(\mu)}\right)} - \int _{-\infty}^{\mu}\ln{\left(1-e^{iS(E)}\right) dE}\biggr),
  \label{eq:2partmag}
  \end{aligned}
\end{equation}
where we use integration by parts to split the integral into two.
The first part of the integral simplifies to:
\begin{equation}
\begin{split}
\Im \frac{d}{dB}\left(\ln{\left(1-e^{iS(\mu)}\right)}\right) = -\Im \left(\frac{i e^{iS(E)}}{1-e^{iS(E)}} \frac{d S(E)}{dB}\right) =
-\Im \left(i\frac{1-e^{-iS(E)}}{2-2\cos{\left(S\right)}} \frac{d S(E)}{dB}\right) = \frac{1}{2} \frac{d S(E)}{dB}.
\end{split}
\label{eq:mag_zero}
\end{equation}
From Eq.~\eqref{eq:action}, we see that the Eq.~\eqref{eq:mag_zero} will average out to zero in an integral over $k_z$ and therefore will not contribute to the magnetisation.
As a result, we focus our attention on the second term in Eq.~\eqref{eq:2partmag}
\begin{equation}
\begin{split}
  M(k_y, \mu) = -\frac{1}{W \pi^2} \int_{-\pi/c}^{\pi/c} dk_z \Im \int_{-\infty}^\mu dE \frac{i e^{iS(E)}}{1-e^{iS(E)}} \frac{d S(E)}{dB}
  \label{eq:real_mag}
\end{split}
\end{equation}
To complete the integral over $E$ in Eq.~\eqref{eq:real_mag}, we employ analytic continuation and extend the energy into the complex plane $E + i\mathcal{E}$.
We use the following integration contour as shown in Fig.~\ref{fig:suppl}:
\begin{enumerate}[I.]
  \item $E \to -\infty$ and $\mathcal{E} \in (0, +\infty)$.
  \item $E \in (-\infty, \mu)$ and $\mathcal{E} \to +\infty$
  \item $E = \mu$ and $\mathcal{E} \in (+\infty, 0)$.
\end{enumerate} 
One can check through Eq.~\eqref{eq:action} that integral contours \upperRomannumeral{1} and \upperRomannumeral{2} do not contribute because $\lim_{E \to -\infty} e^{iS (E)} = 0$ and $\lim_{\mathcal{E} \to \infty} e^{iS (E)} = 0$.
To proceed with the remaining integral \upperRomannumeral{3}, we work with open trajectories which arise when $k_x > 2 t_\perp/(\hbar v_x)$.
We argue that this relation breaks only when the trajectories are aligned with the magnetic field direction and therefore should not contribute to the magnetisation.
We express action in terms of linearised momentum as we did in Eq.~\eqref{eq:action}, but now include a first order $\mathcal{E}$ term:
\begin{equation}
\begin{aligned}  
  S (\mu+i\mathcal{E}, k_y, k_z) = 2 k_F W + i\frac{\mathcal{E}}{\Delta{E_x}} - \frac{2\Gamma \cos{\left(\phi/2+k_z c\right)}}{\Delta E_x},
  \label{eq:action_im}
\end{aligned}
\end{equation}
where we define $k_F = k_x(k_y, \mu)$ and $\Delta{E_x} = \Delta E_x (\mu, k_y)$ and exclude the explicit dependence on $k_y$ and $\mu$ in the equations for brevity.
Higher order $\mathcal{E}$ terms are neglected due to the exponential decay $e^{-\mathcal{E}/\Delta{E_x}}$ of the ingrand.
We substitute Eq.~\eqref{eq:action_im} into the integral Eq.~\eqref{eq:real_mag} along contour \upperRomannumeral{3}:
\begin{equation}
  \begin{aligned}
  M(k_y, \mu) = \frac{\Delta{E_x}}{W \pi^2} \int_{-\pi/c}^{\pi/c} dk_z \Im \int_{0}^{\infty} d\mathcal{E} \frac{i e^{-\frac{\mathcal{E}}{\Delta{E_x}}+iS(\mu)}}{1-e^{-\frac{\mathcal{E}}{\Delta{E_x}}+iS(\mu)}} \frac{d S (\mu)}{dB} = -\frac{\Delta{E_x}}{W \pi^2} \Im  \int_{-\pi/c+\phi/2}^{\pi/c + \phi/2} dk_z i \ln{\left(1-e^{iS(\mu)}\right)} \frac{d S(\mu)}{dB},
  \label{eq:im_mag}
  \end{aligned}
\end{equation}
\begin{equation}
  \begin{aligned}
  \frac{dS}{dB} = \frac{ecW}{\hbar \Delta{E_x}} \left[\Gamma\sin{\left(k_z c\right)} - \frac{d\Gamma}{d \phi} \cos{\left(k_z c\right)}\right].
  \label{eq:action_deriv}
  \end{aligned}
\end{equation}
We expand the natural logarithm $\ln$ in Eq.~\eqref{eq:real_mag} in a power series:
\begin{equation}
\ln{\left(1-e^{iS}\right)} \frac{d S}{dB} = \sum_{p=1}^{\infty} \frac{e^{iSp}}{p} \frac{dS}{dB},
\label{eq:ln_power}
\end{equation}
and utilize the Jacobi-Anger expansion on the exponential term in Eq.~\eqref{eq:ln_power}:
\begin{equation}
  \begin{aligned}
  e^{inS} = e^{i2nk_F W}\biggl[J_0\left(\frac{2n\Gamma}{\Delta{E_x}}\right) +
  2  \sum_{q=1}^{\infty}i^{q} J_q\left(-\frac{2 n\Gamma} {\Delta E_x}\right) \cos{\left(q k_z c\right)}\biggr],
  \label{eq:jacobi-anger}
  \end{aligned}
\end{equation}
where $J_q$ are the Bessel functions of the first kind. Due to the orthogonality between trigonometric function in Eq.~\eqref{eq:action_deriv} and Eq.~\eqref{eq:jacobi-anger}, the integration in Eq.~\eqref{eq:im_mag} will only leave the $q=1$ term remaining:
\begin{equation}
  M(k_y, \mu) = \frac{1}{W c \pi} \frac{d \Gamma}{dB} \sum_{n=1}^{\infty} \frac{\sin{\left(2n k_F W\right)}}{n} J_1\left(\frac{2n\Gamma} {\Delta{E_x}}\right)
  \label{eq:mag_full_ap}.
\end{equation}

\section{Effect of Temperature}

Equation~\eqref{eq:3D_integral} is a result at zero temperature.
The addition of temperature can simplify the form of magnetisation by removing the higher frequency $n>1$ components.
To achieve this, we write $\mathcal{M}$ at non-zero temperature:

\begin{equation}
  \begin{aligned}
\mathcal{M}(\mu, T) = \int_{-\infty}^{\infty} \mathcal{M}(E, T=0) \frac{d F\left(E-\mu, T\right)}{dE} dE  =\left( \mathcal{M} \circledast \frac{dF}{dE} \right) (\mu) = \mathcal{F}^{-1}\left\{\mathcal{F} \left\{\mathcal{M}\right\} \cdot \mathcal{F}  \left\{\frac{dF}{dE}\right\}  \right\}
  \label{eq:mag_temp},
  \end{aligned}
\end{equation}
where $F(E-\mu, T)$ is the Fermi-Dirac distribution. 
In the last part of the equation, we utilize the convolution theorem to express magnetisation as a product of Fourier components of zero temperature magnetisation and Fermi-Dirac distribution.
The Fourier Transform of a Fermi-Dirac distribution is~\cite{fourier_FD}:
\begin{equation}
  \begin{aligned}
    \mathcal{F}  \left\{F \right\} (s) = \frac{1}{\sqrt{2\pi }}\int_{-\infty}^{\infty} F(E) e^{-isE} dE =
    \sqrt{\frac{\pi}{2}} \delta(s) + \frac{1}{\sqrt{2\pi}} \left(\frac{\sinh{(s \pi k_B T)}}{ i \pi k_B T}\right)^{-1}
   \label{eq:FD_fourer}.
  \end{aligned}
\end{equation}
On the other hand, to find the Fourier transform, we work with the linearised momentum:
\begin{equation}
  k_x(E) = K_F + \frac{1}{\Delta{E_x} W} \left(E-\mu\right),
  \label{eq:lin_mom}
\end{equation}
where we drop the $k_y$ index since $k_x(E)$ is evaluated at the extrema.
The zero temperature magnetisation at energy $E$ close to Fermi level $\mu$ reads:
\begin{equation}
  \mathcal{M}(E) =\mathcal{M}_0 \frac{d {\rm sinc(\phi/2)}}{d\phi} \sum_{n=1}^{\infty} \frac{\sin{\left(2n W K_F + \frac{n(E-\mu)}{\Delta{E_x} }\right)}}{n^{3/2}} J_1\left(\frac{2\Gamma} {\Delta{E_x}}\right),
  \label{eq:approximate_3D_mag}
\end{equation}
where only the $\sin$ term was substituted in by Eq.~\eqref{eq:lin_mom} and other terms are kept constant $K_F$ as a result of a steepest-descent approximation.
The approximation is valid as long as $\mu \gg k_B T$.
Therefore, the Eq.~\eqref{eq:FD_fourer} becomes:
\begin{equation}
  \begin{aligned}
    \mathcal{M}(\mu) =\mathcal{M}_0 \frac{d {\rm sinc(\phi/2)}}{d\phi} \sum_{n=1}^{\infty} \left(\frac{\sinh{(\frac{n \pi k_B T}{\Delta{E_x}})}}{ \frac{\left(n \pi k_B T\right)}{\Delta{E_x}}}\right)^{-1} \frac{\sin{\left(2n W K_F\right)}}{n^{3/2}} J_1\left(\frac{2n\Gamma} {\Delta{E_x}}\right).
  \label{eq:mag_temp_complete}
  \end{aligned}
\end{equation}
Whenever the thermal temperature is comparable to or larger than the energy spacing, $k_B T \approx \Delta{E_x}$, the Eq.~\eqref{eq:mag_temp_complete} reduces to a single component:
\begin{equation}
  \mathcal{M}(\mu) \approx \mathcal{M}_0 \frac{d {\rm sinc(\phi/2)}}{d\phi} \sin{\left(2 K_F W \right)} J_1\left(\frac{2\Gamma} {\Delta{E_x}}\right).
  \label{eq:mag_temp_complete_broadened}
\end{equation}


\end{document}